\documentclass[journal]{IEEEtran}

\usepackage{amsmath}
\usepackage{amsfonts}
\usepackage{amssymb}
\usepackage{amsthm}
\usepackage{tabularx}
\usepackage{mathrsfs} 

\usepackage{url}
\usepackage{hyperref}

\usepackage{stfloats}
\usepackage{float}
\usepackage{graphicx}
\usepackage{cite}
\usepackage{xcolor}
\usepackage{subfigure}

\makeatletter
\def\blfootnote{\xdef\@thefnmark{}\@footnotetext}
\makeatother

\begin{document}
	
		\title{\LARGE{Covert Communications over Enormous Fluid Antenna Systems (E-FAS): Impact of Channel Hardening on Detection}} 
	\author{Farshad~Rostami~Ghadi$^\dagger$, Damoon Shahbaztabar$^\ddagger$, Kai-Kit Wong$^\dagger$, and Wei-Ping Zhu$^\ddagger$\\
		$^\dagger$ Department of Electronic and Electrical Engineering, University College London, UK\\
	$\ddagger$	Department of Electrical and Computer Engineering, Concordia University, Canada\\
	(E-mail:$\{\rm f.rostamighadi, kai-kit.wong\}@ucl.ac.uk$, $\rm weiping@ece.concordia.ca$, $\rm damoon.shahbaztabar@mail.concordia.ca$)}
	\maketitle
\begin{abstract}
	Enormous fluid antenna systems (E-FAS) enable guided surface wave (SW) propagation and induce an end-to-end channel that is conditionally complex Gaussian with a random covariance scale. This paper investigates the impact of this channel structure on covert communication in the presence of a passive warden. Under an equal power effective mode representation, the random channel scale follows a Gamma distribution, leading to a Bessel-$K$ distribution for the effective channel power. We show that the optimal likelihood ratio test at the warden reduces to an energy detector and derive the corresponding false alarm and missed detection probabilities. These results are then used to characterize the achievable covert throughput under a prescribed detection error constraint. Numerical results show that, for finite observation intervals, finite mode E-FAS channels improve covertness relative to the fully hardened Rayleigh limit with the same average channel power. They further reveal an asymmetric role of channel hardening: stronger hardening can improve the legitimate link, whereas retaining channel scale fluctuations at the warden can enhance finite observation covertness.
\end{abstract}

	\maketitle
	
	\vspace{0mm}
\section{Introduction}

Covert communication aims to conceal the existence of a wireless transmission from a passive warden, commonly referred to as Willie. Therefore, the problem is  fundamentally one of statistical detection, in which Willie attempts to distinguish between observations obtained in the absence and presence of transmission \cite{Bash2013,Chen2023}. Existing approaches have considered artificial noise, noise uncertainty, friendly jamming, and propagation reconfiguration \cite{He2017,Sobers2017}. More recently, reconfigurable intelligent surfaces (RIS) and fluid antenna systems (FAS) have been investigated as additional means of controlling the signal observed by the warden \cite{Wu2022,Wong2021,ghadi2025}.

Enormous fluid antenna systems (E-FAS) provide a different propagation mechanism \cite{wong2025}. In E-FAS, intelligent surfaces can transport guided surface waves (SWs) along structures such as walls, ceilings, and facades before reradiating the signal toward the intended receiver \cite{Ghadi2026EFAS}. The resulting end to end channel is conditionally complex Gaussian with a random covariance scale, while its unconditional distribution is generally a Gaussian scale mixture \cite{Ghadi2026EFAS}. The fluctuations of this scale are governed by the effective number of contributing radiation modes. This creates an interesting covert communication tradeoff: channel hardening is beneficial for the intended link, but reducing the channel scale fluctuations at Willie may make the transmitted signal statistically easier to detect.

Motivated by this observation, we study finite observation covert communication over the compound Gaussian E-FAS channel. Under an equal power effective mode representation, we derive the distribution of the effective channel power and show that Willie’s optimum likelihood ratio test reduces to an energy detector. We then characterize the false alarm and missed detection probabilities, the legitimate link outage probability, and the achievable covert throughput. The results reveal an asymmetric effect of E-FAS channel hardening: a large effective mode number can improve reliability at Bob, whereas retaining scale fluctuations at Willie can improve finite observation covertness.

\section{System Model}

We consider an E-FAS-assisted downlink consisting of an $M$-antenna transmitter (Alice), a single-antenna legitimate receiver (Bob), and a passive single-antenna warden (Willie). Alice communicates with Bob through a configured E-FAS route, while Willie observes unintended radiation from the same configured surface aperture. The direct Alice-to-terminal links are assumed to be severely blocked so that the E-FAS-assisted component dominates.

\subsection{E-FAS Channel Model}
The BS-to-surface excitation channel is written as \cite{Ghadi2026EFAS}
\begin{align}
	\mathbf H_{\mathrm{BS-sur}}
	=
	\sqrt{\beta_{\mathrm{BS}}}
	\mathbf R_s^{1/2}
	\mathbf G_{\mathrm{BS-sur}}
	\mathbf R_B^{1/2},
	\label{eq:HBS}
\end{align}
where $\mathbf G_{\mathrm{BS-sur}}$ contains independent $\mathcal{CN}(0,1)$ entries, $\beta_{\mathrm{BS}}$ denotes the large-scale BS-to-surface gain, and $\mathbf R_s$ and $\mathbf R_B$ are the spatial correlation matrices associated with the surface excitation ports and BS antennas, respectively.

We define  $
	\mathbf A
	\triangleq
	\mathbf W_{\mathrm{rad}}\mathbf H_{\mathrm{sur}} $
as the calibrated transfer matrix of the E-FAS route configured for Bob, where $\mathbf H_{\mathrm{sur}}$ describes the guided SW propagation and $\mathbf W_{\mathrm{rad}}$ represents the corresponding surface re-radiation mapping. Hence, the radiation-to-terminal channel for $u\in\{b,w\}$ is modeled as
\begin{align}
	\mathbf h_{\mathrm{rad},u}
	=
	\sqrt{q_u}\,
	\mathbf g_u\mathbf R_{L,u}^{1/2},
	\label{eq:hrad}
\end{align}
where $\mathbf g_u\sim\mathcal{CN}(\mathbf 0,\mathbf I)$, $\mathbf R_{L,u}$ denotes the correlation matrix across the effective radiation modes, and
$
	q_u
	\triangleq
	\eta_{L,u}G_{\mathrm{rad},u}\ell_{\mathrm{RU},u}$.  Here, $\eta_{L,u}$ is the guided-to-radiated conversion efficiency, $G_{\mathrm{rad},u}$ denotes the radiation gain in the direction of terminal $u$, and $\ell_{\mathrm{RU},u}$ captures the relative distance dependence of the final space wave segment. In particular, $G_{\mathrm{rad},w}$ does not represent a route intentionally configured toward Willie; it characterizes the unintended radiation observed in Willie's direction. Therefore, the resulting E-FAS-assisted vector channel is defined as
\begin{align}
	\mathbf t_u
	=
	\mathbf h_{\mathrm{rad},u}
	\mathbf A
	\mathbf H_{\mathrm{BS-sur}}.
	\label{eq:tu}
\end{align}

We assume that Alice employs a deterministic unit-norm statistical beamforming vector $\mathbf w$, independent of the instantaneous small-scale fading.  Thus, the corresponding scalar channel is  $ h_u = \mathbf t_u\mathbf w$. 

We now define
$
	\delta
	\triangleq
	\mathbf w^H\mathbf R_B\mathbf w
$
and
$
	\mathbf B
	\triangleq
	\mathbf A\mathbf R_s\mathbf A^H
$. 
So, conditioned on $\mathbf h_{\mathrm{rad},u}$, \eqref{eq:HBS} and \eqref{eq:tu} yield
\begin{equation}
	h_u
	\mid
	\mathbf h_{\mathrm{rad},u}
	\sim
	\mathcal{CN}(0,V_u),
	\label{eq:conditional_h}
\end{equation}
where
\begin{equation}
	V_u
	=
	\delta\beta_{\mathrm{BS}}
	\mathbf h_{\mathrm{rad},u}
	\mathbf B
	\mathbf h_{\mathrm{rad},u}^{H}.
	\label{eq:Vu}
\end{equation}

Then, substituting \eqref{eq:hrad} into \eqref{eq:Vu} gives
\begin{align}
	V_u
	=
	\delta\beta_{\mathrm{BS}}q_u
	\mathbf g_u\mathbf C_u\mathbf g_u^H,
	\label{eq:Vu_quad}
\end{align}
where
$
	\mathbf C_u
	\triangleq
	\mathbf R_{L,u}^{1/2}
	\mathbf B
	\mathbf R_{L,u}^{1/2}$.  Therefore, the average channel scale is obtained as
\begin{equation}
	\Omega_u
	\triangleq
	\mathbb E\{V_u\}
	=
	\delta\beta_{\mathrm{BS}}q_u
	\operatorname{tr}(\mathbf C_u).
	\label{eq:Omega}
\end{equation}
Following the E-FAS channel hardening characterization, the effective number of contributing modes is
\begin{align}
	m_u
	\triangleq
	N_{\mathrm{eff},u}
	=
	\frac{\operatorname{tr}^2(\mathbf C_u)}
	{\operatorname{tr}(\mathbf C_u^2)}.
	\label{eq:Neff}
\end{align}

\subsection{Effective Mode Representation}
For analytical tractability, we first consider the case where the $m_u$ nonzero eigenvalues of $\mathbf C_u$ are equal. Then, \eqref{eq:Vu_quad} becomes the sum of $m_u$ independent exponentially distributed contributions, thereby
\begin{equation}
	V_u
	\sim
	\mathrm{Gamma}
	\left(
	m_u,\frac{\Omega_u}{m_u}
	\right),
	\label{eq:Vgamma}
\end{equation}
where the second argument denotes the scale parameter. For a general eigenvalue profile, \eqref{eq:Vgamma} can be used as a moment matching approximation by defining $m_u$ according to \eqref{eq:Neff}; it preserves both $\mathbb E\{V_u\}$ and $\mathrm{Var}(V_u)$.

Now, we define the instantaneous effective channel power as
$
	Z_u
	\triangleq
	|h_u|^2$. 
Conditioned on $V_u=v$, \eqref{eq:conditional_h} gives
\begin{align}
	f_{Z_u|V_u}(z|v)
	=
	\frac{1}{v}\exp\left(-\frac{z}{v}\right),
	\qquad z\geq 0.
\end{align}
By averaging with respect to \eqref{eq:Vgamma}, we have
\begin{align}
	f_{Z_u}(z)
	=
	\frac{2}{\Gamma(m_u)\theta_u}
	\left(
	\frac{z}{\theta_u}
	\right)^{\frac{m_u-1}{2}}
	K_{m_u-1}
	\left(
	2\sqrt{\frac{z}{\theta_u}}
	\right),
	\label{eq:fZ}
\end{align}
where
$
	\theta_u
	\triangleq
	\frac{\Omega_u}{m_u}$, 
and $K_{\nu}(\cdot)$ denotes the modified Bessel function of the second kind. Furthermore, the corresponding complementary CDF can be derived as 
\begin{equation}
	\overline F_{Z_u}(z)
	=
	\frac{2}{\Gamma(m_u)}
	\left(
	\frac{z}{\theta_u}
	\right)^{m_u/2}
	K_{m_u}
	\left(
	2\sqrt{\frac{z}{\theta_u}}
	\right).
	\label{eq:ccdfZ}
\end{equation}
Consequently, the first two moments are
$
	\mathbb E\{Z_u\}
	=
	\Omega_u$
and
$
	\mathrm{Var}(Z_u)
	=
	\Omega_u^2
	\left(
	1+\frac{2}{m_u}
	\right)
$. As $m_u\rightarrow\infty$, the random scale concentrates around its mean,
$
	V_u
	\xrightarrow{\mathrm{p}}
	\Omega_u$
and thus
\begin{equation}
	Z_u
	\rightarrow
	\mathrm{Exp}(\Omega_u).
	\label{eq:hardlimit}
\end{equation}
Hence, the conventional Rayleigh channel is recovered as the fully hardened E-FAS limit.

\subsection{Covert Transmission Model}
We assume that Alice transmits $N$ independent Gaussian symbols $x[n]\sim\mathcal{CN}(0,1)$ with power $P$. The channels are assumed to remain constant over the $N$ symbol observation interval and vary independently across transmission blocks. Therefore, the received signal at Bob is defined as
\begin{align}
	y_b[n]
	=
	\sqrt{P}h_bx[n]+n_b[n],
	\label{eq:yb}
\end{align}
where $n_b[n]\sim\mathcal{CN}(0,\sigma_b^2)$. Moreover, Willie tests between the hypotheses $	\mathcal H_0$ and $	\mathcal H_1$ as follows
\begin{align}
	\mathcal H_0:\quad
	y_w[n]
	&=
	n_w[n],
	\label{eq:H0}
	\\
	\mathcal H_1:\quad
	y_w[n]
	&=
	\sqrt{P}h_wx[n]+n_w[n],
	\label{eq:H1}
\end{align}
for $n=1,\ldots,N$, where $n_w[n]\sim\mathcal{CN}(0,\sigma_w^2)$.

We assume that Willie knows $P$, $\sigma_w^2$, $N$, and the statistical distribution of $Z_w$, but does not know its instantaneous realization. Thus, Willie is not assumed to have instantaneous E-FAS channel state information. 
Importantly, we assume that $\sigma_w^2$ is  perfectly known. Therefore, any covertness advantage obtained below originates from the E-FAS propagation statistics rather than from an assumed uncertainty in Willie's receiver noise power.

\section{Covert Detection and Performance Analysis}

\subsection{Optimal Detection at Willie}

Here, we define 
$
	\mathbf y_w
	=
	[y_w[1],\ldots,y_w[N]]^T$. 
Under $\mathcal H_0$, we have 
$
	\mathbf y_w|\mathcal H_0
	\sim
	\mathcal{CN}
	\left(
	\mathbf 0,
	\sigma_w^2\mathbf I_N
	\right)
$.  Now, conditioned on $Z_w=z$, the Gaussian signaling assumption gives
\begin{align}
	\mathbf y_w
	\mid
	Z_w=z,\mathcal H_1
	\sim
	\mathcal{CN}
	\left(
	\mathbf 0,
	(\sigma_w^2+Pz)\mathbf I_N
	\right).
	\label{eq:yH1}
\end{align}

After integrating out $Z_w$, the likelihood ratio can be written as
\begin{align}\nonumber
	\Lambda(\mathbf y_w)
&	=
	\frac{
		p(\mathbf y_w|\mathcal H_1)
	}{
		p(\mathbf y_w|\mathcal H_0)
	}
	=
	\int_{0}^{\infty}
	\left(
	\frac{\sigma_w^2}
	{\sigma_w^2+Pz}
	\right)^N\\
&	\times
	\exp
	\left[
	\|\mathbf y_w\|^2
	\left(
	\frac{1}{\sigma_w^2}
	-
	\frac{1}{\sigma_w^2+Pz}
	\right)
	\right]
	f_{Z_w}(z)\,dz.
	\label{eq:LR}
\end{align}
For every $z>0$, the integrand in \eqref{eq:LR} is strictly increasing in $\|\mathbf y_w\|^2$. Since $f_{Z_w}(z)$ is nonnegative, $\Lambda(\mathbf y_w)$ is also strictly increasing in the received energy. Hence, the optimal likelihood ratio test is equivalent to comparing
\begin{align}
	T_w
	=
	\frac{1}{N}
	\sum_{n=1}^{N}|y_w[n]|^2
	\label{eq:Tw}
\end{align}
with a scalar threshold $\tau$. Willie decides $\mathcal H_1$ when $T_w>\tau$.
\subsection{Detection Error Probability}
Under $\mathcal H_0$, the statistic in \eqref{eq:Tw} satisfies
$
	\frac{NT_w}{\sigma_w^2}
	\sim
	\mathrm{Gamma}(N,1)$. 
Hence, the false alarm probability is derived as
\begin{align}
	P_{\mathrm{FA}}(\tau)
	=
	Q
	\left(
	N,
	\frac{N\tau}{\sigma_w^2}
	\right),
	\label{eq:PFA}
\end{align}
where $Q(a,x)$ denotes the regularized upper incomplete Gamma function.

Under $\mathcal H_1$ and conditioned on $Z_w=z$, we have 
$
	\frac{NT_w}
	{\sigma_w^2+Pz}
	\sim
	\mathrm{Gamma}(N,1)
$. 
Therefore,
\begin{align}
	P_{\mathrm{MD}}(\tau)
	=
	\int_0^\infty
	P
	\left(
	N,
	\frac{N\tau}
	{\sigma_w^2+Pz}
	\right)
	f_{Z_w}(z)\,dz,
	\label{eq:PMD}
\end{align}
where $P(a,x)$ is the regularized lower incomplete Gamma function.

Now, we define the total detection error probability as
\begin{align}
	\xi(\tau,P)
	=
	P_{\mathrm{FA}}(\tau)
	+
	P_{\mathrm{MD}}(\tau).
	\label{eq:xi}
\end{align}
For equal prior probabilities, Willie's optimum detection performance is characterized by
\begin{align}
	\xi^\star(P)
	=
	\min_{\tau\geq0}
	\xi(\tau,P).
	\label{eq:xistar}
\end{align}
An $\epsilon$-covert transmission is required to satisfy
\begin{equation}
	\xi^\star(P)
	\geq
	1-\epsilon.
	\label{eq:covertconstraint}
\end{equation}
Also, the optimization in \eqref{eq:xistar} is one-dimensional. Moreover, when the optimum threshold is interior, it satisfies
\begin{align}
	f_{T_w|\mathcal H_1}(\tau^\star)
	=
	f_{T_w|\mathcal H_0}(\tau^\star).
	\label{eq:threshold}
\end{align}

\subsection{Finite-Observation Nature of the E-FAS Gain}
The random E-FAS channel scale does not make a persistent transmission fundamentally undetectable when Willie knows his noise power exactly. For any fixed $P>0$,
\begin{equation}
	\xi^\star(P)
	\rightarrow 0,
	\qquad
	N\rightarrow\infty.
	\label{eq:Ninf}
\end{equation}

Indeed, under $\mathcal H_0$, $T_w$ converges almost surely to $\sigma_w^2$, whereas under $\mathcal H_1$ it converges to $\sigma_w^2+PZ_w$. Since $Z_w>0$ almost surely, the two hypotheses become asymptotically distinguishable. Accordingly, the benefit investigated in this work is a finite observation effect. For finite $N$, the random E-FAS channel scale broadens the distribution under $\mathcal H_1$ and changes its statistical overlap with the noise-only distribution.

\subsection{Legitimate Link Reliability}
We define Bob's instantaneous received SNR as
$
	\gamma_b
	=
	\frac{PZ_b}{\sigma_b^2}$.  For a target spectral efficiency $R$, an outage occurs when
$
	\log_2(1+\gamma_b)<R$. 
Now, by defining
$
	z_0(P,R)
	=
	\frac{(2^R-1)\sigma_b^2}{P}$, 
the outage probability is
\begin{align}
	P_{\mathrm{out}}(P,R)
	=
	F_{Z_b}
	\left(
	z_0(P,R)
	\right).
	\label{eq:Pout}
\end{align}

Now, by using \eqref{eq:ccdfZ}, the corresponding transmission success probability is determined as
\begin{align}
	P_\mathrm{suc}=\frac{2}{\Gamma(m_b)}
	\left(
	\frac{z_0(P,R)}{\theta_b}
	\right)^{m_b/2}
	K_{m_b}
	\left(
	2\sqrt{
		\frac{z_0(P,R)}{\theta_b}
	}
	\right).
	\label{eq:success}
\end{align}

\subsection{Achievable Covert Throughput}

We define the effective covert throughput as the target spectral efficiency multiplied by the probability of successful reception at Bob,
$
	\mathcal T_{\mathrm c}(P,R)
	=
	RP_\mathrm{suc}
$.  Therefore, the maximum achievable covert throughput is  characterized by
\begin{align}
	\mathcal T_{\mathrm c}^{\star}
	=
	\max_{P,R}\quad
	&
	\mathcal T_{\mathrm c}(P,R)
	\label{eq:optimization}
	\\
	\mathrm{s.t.}\quad
	&
	\xi^\star(P)\geq1-\epsilon,
	\nonumber\\
	&
	0\leq P\leq P_{\max},
	\qquad
	R\geq0.
	\nonumber
\end{align}
The optimization involves only two scalar variables. For a given transmit power, the optimal rate follows from a one-dimensional search over $R$, while the allowable power range is determined by the covert constraint in \eqref{eq:covertconstraint}.

\subsection{Role of E-FAS Channel Hardening}
Regarding the definition of $	\mathrm{Var}(Z_u)$, the normalized variance for a fixed average channel power $\Omega_u$ is derived as
\begin{align}
	\frac{\mathrm{Var}(Z_u)}
	{\mathbb E^2\{Z_u\}}
	=
	1+\frac{2}{m_u}.
	\label{eq:normvar}
\end{align}
Hence, a smaller effective mode number produces stronger received power fluctuations, whereas increasing $m_u$ removes the additional scale randomness and approaches the Rayleigh limit in \eqref{eq:hardlimit}. The roles of this hardening can differ at Bob and Willie. Increasing $m_b$ suppresses the additional channel scale fluctuations experienced by the intended receiver and can improve reliability. However, at Willie, increasing $m_w$ reduces the uncertainty in the received signal strength under $\mathcal H_1$, which can facilitate statistical detection. Accordingly, the covert behavior of E-FAS cannot in general be inferred from its conventional reliability behavior. These results suggest that E-FAS configurations producing a larger effective mode number toward Bob while retaining fewer effective modes in unintended observation directions may be beneficial for covert transmission. The resulting performance behavior is quantified numerically in the next section. \vspace{-4mm}
\section{Numerical Results}
In this section, we evaluate the proposed analytical framework and examine the impact of the E-FAS effective mode numbers on covert detection and throughput. Unless otherwise stated, we set the observation length to $N=20$, the covertness parameter to $\epsilon=0.1$, and normalize the receiver noise powers as $\sigma_b^2=\sigma_w^2=1$. To isolate the effect of the compound Gaussian channel statistics, the average Willie channel power is normalized to $\Omega_w=1$, such that the transmit power can be equivalently represented by the average Willie SNR $\bar{\gamma}_w=P\Omega_w/\sigma_w^2$. For the throughput evaluation, we set $\Omega_b/\Omega_w=10$ dB and optimize the target rate for each operating point, while the transmit power is limited by the covertness constraint; $P_{\max}$ is assumed sufficiently large so that it is not active in the reported results. 

\begin{figure*}[!t]
	\centering
	\includegraphics[width=2\columnwidth]{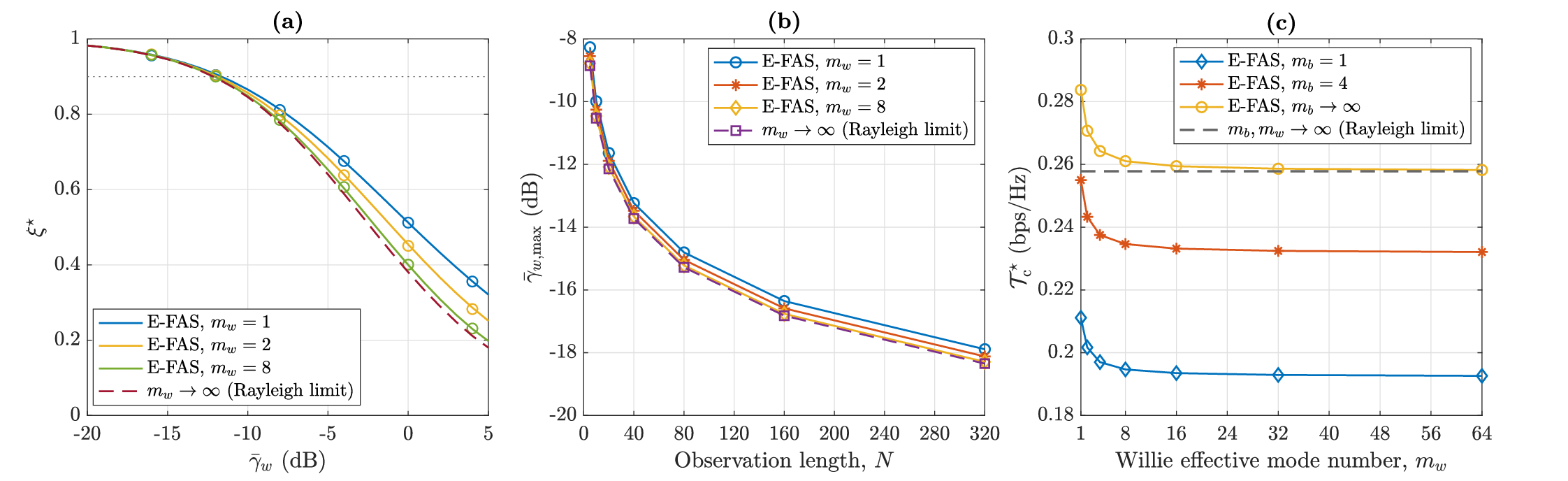}
	\caption{Impact of the E-FAS effective mode number on covert performance: (a) minimum detection error versus average Willie SNR, (b) maximum allowable Willie SNR versus observation length, and (c) optimized covert throughput versus Willie effective mode number for different Bob mode numbers.}\vspace{-4mm}
	\label{fig:2}
\end{figure*}

For Figs. \ref{fig:2}(a) and \ref{fig:2}(b), we consider
$m_w\in\{1,2,8,\infty\}$ to illustrate the transition from
strong channel fluctuations to the Rayleigh limit. In Fig. \ref{fig:2}(c),
$m_w$ is varied from $1$ to $64$, while
$m_b\in\{1,4,\infty\}$ is considered to examine the corresponding
hardening effect on the legitimate link.

Monte Carlo markers are obtained from $10^5$ independent channel and noise realizations. Unless explicitly varied, all other parameters are kept fixed across the compared cases.

Fig. \ref{fig:2} summarizes the effect of the E-FAS effective mode number on covert detection and throughput. In Fig. \ref{fig:2}(a), the minimum detection error decreases as the average received SNR at Willie increases, as expected. More importantly, for the same average channel power, a smaller \(m_w\) consistently results in a larger \(\xi^\star\). The separation becomes more visible as \(\bar{\gamma}_w\) increases. This behavior follows from the additional fluctuations of the random channel scale when only a few effective modes contribute to the received signal. These fluctuations broaden the distribution under \(\mathcal H_1\) and increase its overlap with the noise only distribution. As \(m_w\) increases, the scale fluctuations are suppressed and the detection performance gradually approaches the Rayleigh limit.

Fig. \ref{fig:2}(b) further shows that the maximum Willie SNR satisfying the covertness constraint decreases with the observation length \(N\). With more observations, Willie can estimate the received energy more accurately and therefore detect progressively weaker transmissions. Nevertheless, for any given finite \(N\), a smaller \(m_w\) allows a slightly larger received SNR while maintaining the same detection error requirement. The gap between the finite mode cases and the Rayleigh limit becomes less significant as \(N\) increases, which is consistent with the finite observation nature of the E-FAS covertness gain.

The corresponding throughput behavior is shown in Fig. \ref{fig:2}(c). For a fixed Bob mode number, the optimized covert throughput decreases as \(m_w\) increases because channel hardening at Willie reduces the uncertainty available to conceal the transmission. In contrast, increasing \(m_b\) improves the throughput by reducing the channel fluctuations experienced by Bob and improving the reliability of the intended link. For the considered operating conditions, higher throughput is obtained when Bob experiences a large effective mode number while Willie observes a smaller number of effective modes. As \(m_w\) increases, the curves approach their respective hardened limits, while the case \(m_b\rightarrow\infty\) converges to the fully hardened Rayleigh benchmark. These results highlight the asymmetric role of E-FAS channel hardening: hardening is beneficial at the intended receiver, whereas retaining scale fluctuations at the warden can improve finite observation covert performance.
\section{Conclusion}

We studied covert communication over the compound Gaussian channel induced by E-FAS propagation. Using an effective mode representation, we characterized the random channel scale and the resulting Bessel-$K$ distribution of the received channel power. We showed that the optimal detector at a noncoherent warden reduces to an energy detector and derived the corresponding detection error probabilities, legitimate link outage probability, and achievable covert throughput. The results reveal an asymmetric role of channel hardening: a larger effective mode number can improve reliability at Bob, whereas retaining channel scale fluctuations at Willie can improve finite observation covertness. These results suggest that covert E-FAS design may benefit from jointly accounting for the legitimate link quality and the effective modal structure observed in unintended directions.

\end{document}